\documentclass[12pt]{article}
\usepackage{a4wide}
\usepackage{amssymb,amsmath}
\usepackage{graphicx}
\begin{document}
{\renewcommand{\thefootnote}{\fnsymbol{footnote}}
\begin{center}
{\LARGE Reduction of primordial chaos\\ by generic quantum effects }\\
\vspace{1.5em}
Martin Bojowald,$^1$\footnote{email address: {\tt bojowald@psu.edu}}
David Brizuela,$^2$\footnote{email address: {\tt david.brizuela@ehu.eus}}
Paula Calizaya Cabrera$^3$\footnote{email address: {\tt pcaliz1@lsu.edu}}
and Sara F.~Uria$^2$\footnote{email address: {\tt sara.fernandezu@ehu.eus}}
\\
\vspace{0.5em}
$^1$ Institute for Gravitation and the Cosmos,
The Pennsylvania State
University,\\
104 Davey Lab, University Park, PA 16802, USA\\
\vspace{0.5em}
$^2$ Department of Physics and EHU Quantum Center, University of the Basque Country UPV/EHU,
Barrio Sarriena s/n, 48940 Leioa, Spain\\
\vspace{0.5em}
$^3$ Department of Physics \& Astronomy, Louisiana State University,\\
Baton Rouge, LA 70803, USA\\
\vspace{1.5em}
\end{center}
}

\setcounter{footnote}{0}

\begin{abstract}
  According to general relativity, the generic early-universe dynamics is
  chaotic. Various quantum-gravity effects have been suggested that may change
  this behavior in different ways. Here, it is shown how key mathematical
  properties of the classical dynamics can be extended to evolving quantum
  states using quasiclassical methods, making it possible to apply the
  established dynamical-systems approach to chaos even to quantum
  evolution. As a result, it is found that quantum fluctuations contribute to the reduction of
  the primordial chaos in early-universe models.
\end{abstract}

A powerful statement about the complicated nature of the primordial state of
the universe is made by generic features of chaotic dynamics in classical
descriptions based on general relativity \cite{BKL,Mixmaster}. The evolving
anisotropy of space in the early, high-curvature universe can be described by
an effective potential that encodes the effects of space-time curvature by
walls that restrict the anisotropy parameters to a finite region. Mathematical
results about billiard dynamics applied to these walls, which happen to be
convex and therefore defocusing, then guarantee chaos \cite{Sinai}. Quantum
effects, such as fluctuations or various geometrical implications of
approaches to quantum gravity, may be expected to make this behavior even more
counter-intuitive and harder to unravel. Finding reliable knowledge of the
initial state of the universe could then be impossible.

In particular, a long series of studies in supergravity and string theory has
confirmed this expectation to some degree, showing that the dynamics remains
chaotic \cite{StringChaos,Billiards} when extra dimensions and fields relevant
for unification are included. Such new ingredients extend the classical
configuration space of anisotropy parameters for the universe by including new
independent degrees of freedom. Nevertheless, they come along with their own
curvature contributions that share qualitative features of the walls in the
effective anisotropy potential, maintaining the classical chaotic dynamics.
These models, however, are not fully quantum because they do not consider
states with fluctuations and correlations obeying uncertainty relations.

Independently, quantum cosmology with fluctuating states has also been applied
to this question, but so far with mixed results
\cite{BraneNonChaos,NonChaos,AffineSmooth,ClassQuantMixmaster}, owing for instance to the
difficult task of evaluating dynamical properties of the quantum wave function
of a single universe. The large number of degrees of freedom contained in a
quantum wave function, compared with the corresponding classical degrees of
freedom, makes it hard to disentangle different, potentially competing,
dynamical features. Specific properties are therefore identified in these
approaches that may reduce chaos, such as by isotropization or bounded
curvature. These effects may help to avoid the strong anisotropy potential that
implies chaos in the classical dynamics, but they do not directly confront it.

Here, we apply a systematic quasiclassical expansion to quantum cosmology and
provide the first generic example of a suppression of chaos in primordial
cosmology, based on general features of quantum fluctuations. This new result
relies on a minimal inclusion of quantum effects that implement the spread-out
nature of quantum states along classical trajectories. It should therefore be
present in any model of quantum gravity because it does not require specific
assumptions about the nature of quantum space-time.

To this end, we derive a new effective potential that extends the classical
anisotropy potential into a higher-dimensional quasiclassical configuration
space in which the classical anisotropy parameters are accompanied by
fluctuation degrees of freedom that also parameterize higher moments of an
evolving quantum state. This description makes it possible to retain the
appealing geometrical picture of billiard models \cite{Billiards}, also
extended by new degrees of freedom, but with a specific modification of the
classical walls that has not been considered before. (Related methods have
recently been applied to individual reflections in the anisotropy potential
\cite{QuasiClassKasner}.) At the same time, established methods from
dynamical-systems theory can be applied in the quasiclassical description in order to
analyze detailed quantitative properties of the primordial chaos. While chaos
is not completely removed, its strength is noticeably reduced.

Following the Belinskii--Khalatnikov--Lifshitz (BKL) scenario \cite{BKL}, the
high-curvature dynamics close to the big bang can be studied by analyzing a
spatially homogeneous space-time model with line element
\begin{equation} \label{metric}
  {\rm d}s^2=-N^2(t){\rm d}t^2+\sum_{i=1}^3a_ {i}^2(t)\sigma_i^2,
\end{equation}
where $N(t)$ is the lapse function, $\sigma_i$, with $i=1,2,3$, form a basis
of differential forms on the manifold of the rotation group ${\rm SO}(3)$,
parameterized for instance by Euler angles, and $a_i(t)$ are three independent
functions of time. As the $a_i(t)$ change at different rates, the anisotropy
of the universe evolves, while space expands if the product
$a_1(t)a_2(t)a_3(t)$ increases. The approach to the big-bang singularity can
be studied by inverting the direction of time.

\begin{figure}[h]
  \begin{center}
  \includegraphics[width=0.7\textwidth]{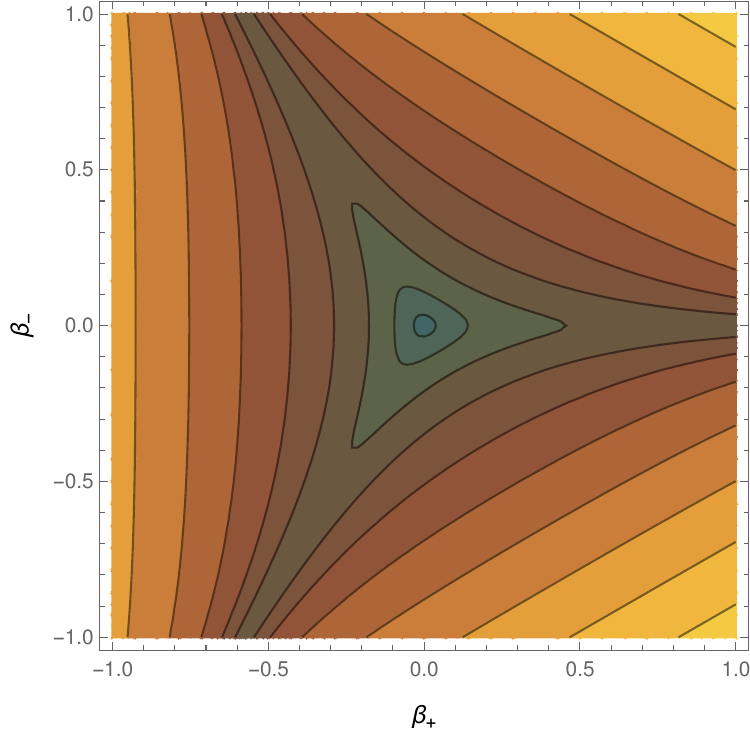}
  \caption{Contour plot of the classical anisotropy potential
    (\ref{potential_Bianchi_IX}). \label{fig:equipotential}}
  \end{center}
\end{figure}

It is useful to separate the changing size of space from the evolution of its
anisotropy, which can conveniently be achieved by introducing Misner variables
\cite{Mixmaster}: The expansion rate is described logarithmically by a real
parameter
\begin{equation}
  \alpha=\frac{1}{3}\ln(a_1a_2a_3)\,,
\end{equation}
such that the big-bang singularity is approached for $\alpha\to-\infty$. The two
remaining degrees of freedom independent of $\alpha$ are the anisotropy
parameters
\begin{equation} \label{beta}
  \beta_+ = \frac{1}{6}\ln(a_1a_2/a_3^2)\quad\mbox{and}\quad
  \beta_- = \frac{1}{2\sqrt{3}} \ln(a_1/a_2)\,.
\end{equation}

General relativity implies that the dynamics of $\alpha$ and $\beta_{\pm}$ in
a universe with line element (\ref{metric}) is described by the expression
\begin{equation}\label{constraint}
{\cal C}=\frac{1}{2} e^{-3 \alpha}
\left(-p_\alpha^2+p_{-}^2+p_{+}^2\right)+e^{\alpha} V(\beta_+,\beta_-)=0,
\end{equation}
with the anisotropy potential
\begin{equation}\label{potential_Bianchi_IX}
V(\beta_+,\beta_-):=\frac{1}{6}\left[
e^{-8\beta_+}+2e^{4\beta_+}\left(
\cosh(4\sqrt{3}\beta_-)-1
\right)-4e^{-2\beta_+}\cosh(2\sqrt{3}\beta_-)
\right]\,.
\end{equation}
(The simple coefficients in the quadratic momentum dependence are a
consequence of the choice of defining (\ref{beta}).)

The expression ${\cal C}$ defined in (\ref{constraint}) serves as a
Hamiltonian constraint: It imposes an energy-balance constraint by being
required to vanish, and, at the same time, it generates Hamilton's equations
for $\alpha$, $\beta_{\pm}$, and their momenta
$p_{\alpha}:=-\frac{e^{3\alpha}}{N} \frac{{\rm d}\alpha}{{\rm d}t}$ and
$p_{\pm}:=\frac{e^{3\alpha}}{N} \frac{{\rm d}\beta_{\pm}}{{\rm d}t}$. Since $\alpha$ changes
monotonically and we are interested in the dynamics of $\beta_{\pm}$ during
expansion or contraction, we can solve the constraint equation
(\ref{constraint}) for
\begin{equation} \label{H}
  H:=-p_\alpha=\sqrt{p_+^2+p_-^2+2e^{4\alpha}V(\beta_+,\beta_-)},
\end{equation}
and view it as the Hamilton function (no longer constrained to vanish) for
$\beta_{\pm}$ and their momenta evolving with respect to $\alpha$. The
dynamics is then determined by the 2-dimensional motion in the anisotropy
potential \eqref{potential_Bianchi_IX}. It has steep exponential walls for large anisotropy and a triangular
symmetry that can best be seen by looking at the contour plot in
Fig.~\ref{fig:equipotential}.

Note that in (\ref{H}) this potential is multiplied by a factor of $e^{4\alpha}$. The
potential walls therefore shrink close to the singularity, giving room for
larger anisotropy. The dynamics of $\beta_{\pm}$ is characterized by a
sequence of reflections at the potential walls. As shown by
Fig.~\ref{fig:equipotential}, these walls are always convex. General
mathematical results can then be invoked to conclude that the dynamics of
anisotropies is chaotic.

Previous results that suggested a reduction or disappearance of chaos
exploited possible ways to tear down the walls of the anisotropy potential,
for instance by using modified dynamics that isotropizes the evolution or
limits the values of curvature and therefore the height of the walls. A
problematic aspect of such modified dynamics is that it would also undermine
the BKL scenario, which in classical general relativity justifies the
application of a spatially homogeneous geometry close to any (spacelike) singularity.
Our new quantum effect does not tear down the walls, but rather, as we will show, reshapes them by making
them partially concave. Concave and therefore focusing walls may or may not
imply chaos \cite{FocusingChaos}. According to our dedicated analysis of the
new system encountered here, the strength of chaos is reduced by a certain
degree depending on how much concavity is included.

As already commented above, the walls are reshaped by quantum
effects. Heuristically, a wave function on the anisotropy plane with
coordinates $\beta_{\pm}$ is spread out and the corners of the triangular
region are therefore washed out, partially closing them off by concave caps. In
order to make this precise, we have to be able to derive a sufficiently
general effective potential that retains the geometrical picture of the
classical model even while it includes quantum effects. Since the anisotropy
parameters may change rapidly when curvature is large, the quantum potential
should be able to capture non-adiabatic effects, ruling out standard methods
such as low-energy effective potentials or other derivative expansions. As we
will show, the expected effect is described well by quasiclassical methods
based on a parameterization of states by moments.

We will consider only quantum states of the anisotropy parameters and their
momenta, and therefore include the central moments
\begin{align}\label{def_moments}
&\Delta(\beta_+^i\beta_-^jp_+^kp_-^l):=\langle
(\hat{\beta}_+-\beta_+)^i(\hat{\beta}_--\beta_-)^j(\hat{p}_+-p_+)^k
(\hat{p}_--p_-)^l
\rangle_{\text{Weyl}},
\end{align}
in completely symmetric or Weyl ordering. The dynamics of a state described by
the moments is governed by a Hamilton operator quantizing
(\ref{constraint}). Its expectation value $\langle\hat{\cal C}\rangle$, taken
in a state described by (\ref{def_moments}), is a function of the moments, for
which a series expansion can be obtained if the anisotropy potential is first
expanded in a Taylor series. As in (\ref{H}), a square root then defines the
effective Hamiltonian $H_{\rm Q}$ with respect to $\alpha$:
\begin{equation} \label{HQ}
  H_{\rm Q}^2=\langle\hat{H}^2(\hat{\beta}_+,\hat{\beta}_-,
\hat{p}_+,\hat{p}_-)\rangle
=p_+^2+\Delta(p_+^2)+p_-^2+\Delta(p_-^2)+
e^{4\alpha}\sum_{i,j=0}^{+\infty}\frac{1}{i!j!}
\frac{\partial^{i+j} V(\beta_+,\beta_-)}{\partial \beta_+^{i}\partial \beta_-^{j}}
    \Delta(\beta_+^i\beta_-^j)\,.
\end{equation}
To simplify the notation, here and from now on we refer to expectation values by their classical
names, $\beta_{\pm}=\langle\hat{\beta}_{\pm}\rangle$ and $p_{\pm}=\langle\hat{p}_{\pm}\rangle$.

By keeping the moments as independent degrees of freedom, in addition to the
expectation values $\beta_{\pm}$ and $p_{\pm}$, we are able to capture
non-adiabatic effects in which moments may change rapidly compared with the
corresponding classical degrees of freedom. In the anisotropy potential of
relevance here, such rapid changes may happen when a state gets squeezed into
the corners of the triangular potential, precisely where we expect spread-out
wave functions to introduce concave components to the walls. However, keeping
all the moments independent takes us from a 2-dimensional configuration plane
to an infinite-dimensional space. For tractability, we have to find a
compromise in which some moment degrees of freedom are kept independent while
most of them are strictly related to the independent ones. An example is a
Gaussian approximation, in which the expectation values and a fluctuation
parameter $s$ are free, while all other moments are specific functions of $s$.

The quasiclassical formulation is open to studies of different classes of
states, defined through specific relationships between the moments. Our main
interest here is to analyze a succession of reflections at steep walls, which
(unlike, say, tunneling) is expected to preserve the shape of a wave
packet. Moments of a single wave packet, such as Gaussian one in which we have
two independent fluctuation parameters, $s_{\pm}$ for $\beta_{\pm}$, should
therefore be reliable. Anisotropy moments are then defined as
\begin{equation} \label{Delta}
  \Delta(\beta_+^{2n}\beta_-^{2m})=\frac{s_+^{2n}s_-^{2m}}
  {2^n2^m} \frac{(2n)!(2m)!}{n!m!}\,.
\end{equation}
The Hamiltonian also contains the momentum terms $p_{\pm}^2$, which, in an
expectation value, imply fluctuation terms:
$\langle\hat{p}_{\pm}^2\rangle= p_{\pm}^2+ \Delta(p_{\pm}^2)$. In order to
obtain a standard dynamical system, we should express the momentum
fluctuations $\Delta(p_{\pm}^2)$ in terms of momenta $p_{s_{\pm}}$ canonically
conjugate to the anisotropy parameters $s_{\pm}$. Methods from Poisson
geometry, which relate the quantum commutator of operators to a Poisson
bracket of their expectation values \cite{EffAc,Karpacz,EffPotRealize}, show
that momentum fluctuations expressed in terms of fluctuation momenta must be
such that
\begin{equation}
 \Delta(\beta_{\pm}p_{\pm})=s_{\pm}p_{s_\pm}\quad,\quad \Delta(p_{\pm}^2)=p_{s_\pm}^2+\frac{U_{\pm}}{s_{\pm}^2},
\end{equation}
where $U_{\pm}=\Delta(p_{\pm}^2)\Delta(\beta_{\pm}^2)- \Delta(\beta_{\pm}p_{\pm})^2 \geq\hbar^2/4$
determines saturation properties of the state. Similar parameterizations have
been used in a variety of fields \cite{VariationalEffAc,SemiClassChaos,QHDTunneling,CQC}.

To conclude this part of our construction, the quantum anisotropy potential in
the 4-dimensional space with coordinates $\beta_{\pm}$ and $s_{\pm}$ receives
two types of new terms: (i) repulsive potentials $U_{\pm}/s_{\pm}^2$ from the
kinetic energy operator, which prevent the fluctuation parameters from
reaching zero and thereby enforce uncertainty relations, and (ii) a series
expansion in $s_{\pm}$ that follows from replacing \eqref{HQ} with
\eqref{Delta}. In fact, after inserting  the parametrization of the moments \eqref{Delta}, this
series can be summed up explicitly, resulting in the extended anisotropy
potential
\begin{eqnarray} \label{VQ}
  &&  V_{\rm Q}(\beta_+,\beta_-,s_+, s_-)\\[8pt]
  &=&\frac{1}{6}
    \bigg[
        e^{-8\beta_++32s_+^2}
    +2e^{4\beta_++8s_+^2} \left(e^{24s_-^2}\cosh(4\sqrt{3}\beta_-)-1\right)
    -4
    e^{-2\beta_++2s_+^2+6s_-^2}\cosh(2\sqrt{3}\beta_-)
    \bigg]\,. \nonumber
\end{eqnarray}
    
The 4-dimensional nature of the new configuration space complicates our
heuristic arguments because the expected concave caps around the triangular
corners of the classical potential depend on which 2-dimensional cross-section
we consider. Nevertheless, a crucial difference implied by non-zero $s_{\pm}$
can be seen by considering, for instance, the cross-section defined by
$\beta_-=0$. The classical potential is then reduced to the simple function
$6V(\beta_+,0)= e^{-8\beta_+}-4e^{-2\beta_+}$. The two exponentially
decreasing terms illustrate the steep exponential wall on the left of
Fig.~\ref{fig:equipotential}, which approaches zero on the right where the
confined region is stretched out into a narrow channel. The extended
potential, evaluated on the same cross-section, implies the reduced potential
\begin{equation} \label{Wall}
  6V_{\rm Q}(\beta_+,0,s_+, s_-)=
                   e^{-8\beta_++32s_+^2}-4
                                  e^{-2 \beta_++2s_+^2+6s_-^2}+2e^{4\beta_++8s_+^2} \left(e^{24s_-^2}-1\right),
 \end{equation}
 in which an exponentially increasing function contributes as well. Since
 $s_-$ cannot be zero, owing to the $U_-/s_-^2$ term in the Hamiltonian, even
 a small value will eventually lead to the increasing $\exp(4\beta_+)$ to
 dominate for positive $\beta_+$. The classical channel has thus been closed
 off, which requires a concave contribution to the walls.  (In fact, for large
 $s_-$, almost the entire wall would become concave; see App.~\ref{a:Wall}.)
 Actually, this is a generic feature for any quantum state since the operator
 $(\cosh(4\sqrt{3}\hat{\beta}_-)-1)$, which multiplies $\exp(4\hat{\beta}_+)$
 in the potential \eqref{potential_Bianchi_IX}, is positive definite; see
 App.~\ref{a:PosDef}. Therefore, in the effective potential,
 $\langle\exp(4\hat{\beta}_+)(\cosh(4\sqrt{3}\hat{\beta}_-)-1) \rangle$
 produces a dominant exponential contribution for large $\beta_+$, regardless
 of the value of $\beta_-$, which closes the classical channel on the right
 corner.  Similar arguments can be applied to the other two classical channels
 located at $\beta_-=\pm\sqrt{3}\beta_+$.

   \begin{figure}[h]
     \begin{center}
 \includegraphics[width=0.8\textwidth]{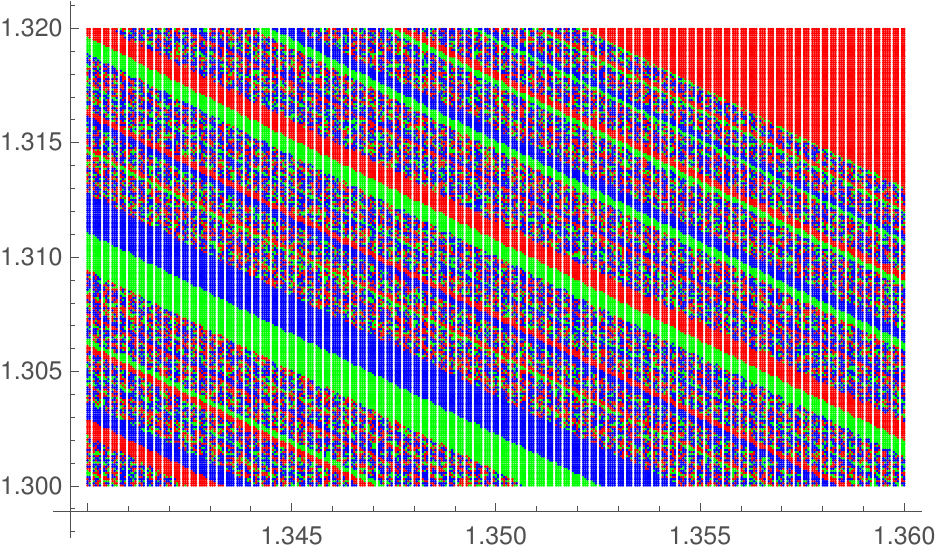}
 \includegraphics[width=0.8\textwidth]{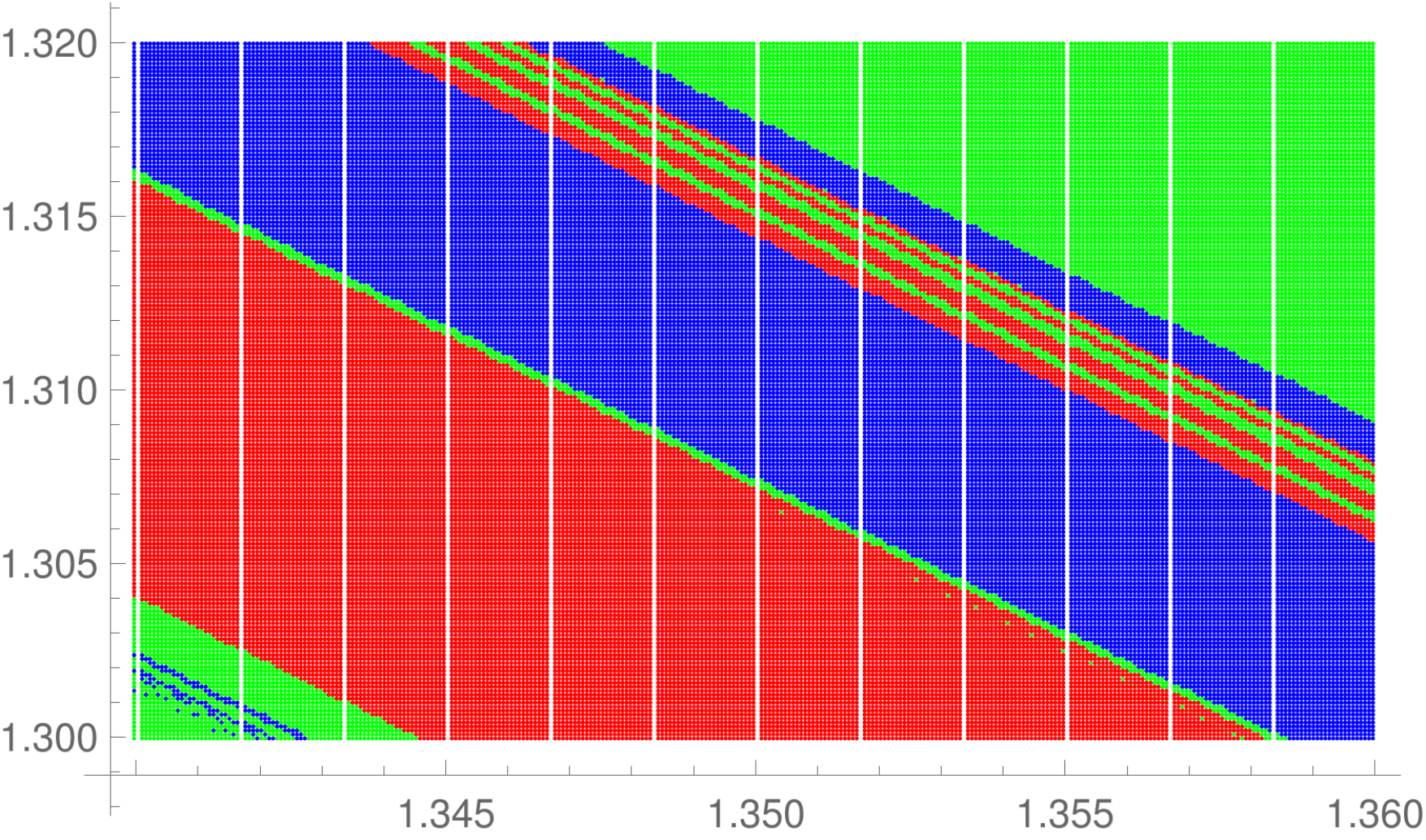}
  \caption{Subsets of three outcomes in a plane of initial values of the anisotropies, classical
    (top) and quasiclassical (bottom). Each color (red, blue, and green)
    stands for the corner of the potential that the corresponding point hits
    first. \label{fig:Fractal}}
  \end{center}
\end{figure}

Since the 4-dimensional dynamics in the extended anisotropy potential
(\ref{VQ}) is given by a dynamical system in canonical form, its properties of
chaos can be unambiguously analyzed. Details of such an analysis are
presented in Ref.~\cite{Long_paper}, and we note that the two main methods that have been used
in the past to derive coordinate-independent properties of chaos agree in that
quasiclassical effects lead to a reduction of its degree. First, there exists
a canonical transformation that maps our 4-dimensional dynamical system to
motion within a finite region of a 4-dimensional space with constant negative
curvature, generalizing a crucial property of the Misner--Chitre
transformation \cite{Chitre} used in \cite{ChaosMisnerChitre} to conclude that
all Lyapunov exponents are positive. The quasiclassical dynamics therefore
remains chaotic.

Secondly, a quantitative measure of chaos can be obtained from the fractal
dimension of subsets of different dynamical outcomes in the space of initial
values \cite{BasinChaos,Farey}. Here, three different outcomes are naturally
defined by which of the three corners in the classical anisotropy plane will
be approached first, a definition that can also be used in the quasiclassical
potential.

In our analysis, we have chosen 160 different sets of initial
data for the quantum variables $(s_\pm,p_\pm, U_\pm)$, and divide the space of initial data in corresponding
outcomes. For our sample, the dimension of the boundary between different regions
has been computed to be in the range $[1.21,1.58]$, depending on
the specific initial values of the quantum variables,
while the classical
repeller has a fractal dimension of $1.86$ (see Ref.~\cite{Long_paper} for more details).
Therefore, in all the analyzed cases, quantum fluctuations decrease
this dimension, which implies a reduction of the chaotic behavior.
This can be clearly seen in Fig.~\ref{fig:Fractal}, where the fractal
picture is smoothed out in the quantum case.
In particular, specific
details of the reduction depend on quantum-information properties of the
primordial state of the universe.

\section*{Acknowledgements}

DB and SFU thank, respectively, the
Max-Planck-Institut f\"ur Gravitationsphysik
and The Pennsylvania State University
for hospitality while part of this work was done.
SFU was funded by an FPU fellowship and a mobility
scholarship of the Spanish Ministry of Universities.
This work was supported in part by NSF grant PHY-2206591,
the Alexander von Humboldt Foundation,
the Basque Government Grant \mbox{IT1628-22}, and
by the Grant PID2021-123226NB-I00 (funded by MCIN/AEI/10.13039/501100011033 and by “ERDF A way of making Europe”).

\begin{appendix}

\section{Turn-over from convex to concave}
\label{a:Wall}

For simplicity, let us analyze the convexity of the wall around the corner to the right of the trapped region
(see Fig. 1). This wall is described by the term
\[
W(\beta_+,\beta_-,s_-)=e^{4\beta_+} \left(e^{24s_-^2}
\cosh(4\sqrt{3}\beta_-)-1\right),
\]
because the other terms in the effective potential (11) decrease exponentially for large $\beta_+$. Convexity
of the wall means that the gradient vector of $W$, which is normal to
constant-$W$ surfaces, has a decreasing angle with the $\beta_+$-axis as
$\beta_-$ increases. A straightforward calculation determines this angle
$\theta$ through
\[
\tan\theta= \sqrt{3} \frac{\sinh(4\sqrt{3}\beta_-)}{\cosh(4\sqrt{3}\beta_-)-
	e^{-24s_-^2}}\,.
\]
Hence, the wall is convex while $\partial\tan\theta/\partial\beta_-<0$. For the classical model ($s_-=0$), this is always the case, which, as mentioned, is an indicator of chaos. However, once that quantum effects are taken into account,
some parts of the wall become concave. In particular, it turns
into concave behavior when
\[
\frac{\partial\tan\theta}{\partial\beta_-}=12
\frac{1-e^{-24s_-^2}\cosh(4\sqrt{3}\beta_-)}{(\cosh(4\sqrt{3}\beta_-)-
	e^{-24s_-^2})^2}=0\,,
\]
at which point we have
\[
\tan\theta= \frac{\sqrt{3}}{\sqrt{1-e^{-48s_-^2}}}\,.
\]
For large $s_-$, this turn-over point takes place at $\tan\theta=\sqrt{3}$, or $\theta=60^{\circ}$, that is, just about the middle of the wall.

\section{State-independent closure of corners}
\label{a:PosDef}

Here we prove that the corners of the classical potential are closed off in any
effective potential, not only for the
quantum states with high-order
moments parameterized by (9).

We first demonstrate the strict inequality
$\langle\cosh(4\sqrt{3}\hat{\beta}_-)\rangle>1$ for any state of our model quantized on the Hilbert space
$L^2({\mathbb R}^2,{\rm d}\beta_+{\rm d}\beta_-)$. Using the
$\beta$-representation of wave functions $\psi(\beta_+,\beta_-)$, we conclude
that the operator $\sinh^2(4\sqrt{3}\hat{\beta}_-)$ is positive definite
because there is no (normalizable) state in which
$\langle\sinh^2(4\sqrt{3}\hat{\beta}_-)\rangle=0$, which can be seen on
the spectral decomposition of $\hat{\beta}_-$: On the spectrum, the operator
$\sinh^2(4\sqrt{3}\hat{\beta}_-)$ takes positive values unless
$\beta_-=0$. Since the spectrum is continuous, there is no normalizable state
solely supported on this value, and therefore
$\langle\sinh^2(4\sqrt{3}\hat{\beta}_-)\rangle$ cannot be zero.

Then, since
$\sinh^2(4\sqrt{3}\hat{\beta}_-)=\cosh^2(4\sqrt{3}\hat{\beta}_-)-1=(\cosh(4\sqrt{3}\hat{\beta}_-)-1)(\cosh(4\sqrt{3}\hat{\beta}_-)+1)$, and  $\cosh(4\sqrt{3}\hat{\beta}_-)+1$ is also positive
definite, $\cosh(4\sqrt{3}\hat{\beta}_-)-1$ must be positive definite as well. Therefore, $\langle\cosh(4\sqrt{3}\hat{\beta}_-)\rangle>1$ in any state.

We now apply this result to a generic effective potential obtained by taking
the expectation value of a quantized (5) in some
family of states, with high-order moments not necessarily
parameterized by (9). Since $e^{4\hat{\beta}_+}$
and $\cosh(4\sqrt{3}\hat{\beta}_-)-1$ are two commuting positive definite
operators, their product is positive definite. Therefore,
$\langle e^{4\hat{\beta}_+}(\cosh(4\sqrt{3}\hat{\beta}_-)-1)\rangle$ cannot be
zero. The corresponding reduced potential, defined as in (12),
then always retains a contribution from the term
$e^{4\hat{\beta}_+}(\cosh(4\sqrt{3}\hat{\beta}_-)-1)$, which is unbounded in
$\beta_+$, and implies an exponential wall for any family of states with
increasing $\beta_+$, irrespective of the value of
${\beta}_-$. This is in contrast with the
classical case, for which this term exactly vanishes at the axis $\beta_-=0$,
and, since the remaining terms in the classical potential decrease with $\beta_+$,
the classical exit channel on the right corner of Fig. 1 is open.

The reduced potential (12) is an
explicit expression obtained for states red with moments parameterized by (9), but its crucial
feature remains valid for any family of states.
By means of symmetry, the same argument can be applied for the remaining two corners $\beta_-=\pm\sqrt{3}\beta_+$.

\end{appendix}

\end{document}